\documentclass[twoside,twocolumn,english]{revtex4-2}
\usepackage[T1]{fontenc}
\usepackage[latin9]{inputenc}
\usepackage{amsmath}
\usepackage{amssymb}
\usepackage{graphicx}
\usepackage{esint}

\makeatletter
\usepackage{soul}
\usepackage{color}
\usepackage{babel}

\makeatother

\begin{document}
\title{Extremal fluctuations driving the relaxation in glassy energy landscapes}
\author{Stefan Boettcher$^{1}$}
\author{Paula A. Gago$^{2}$}
\author{Paolo Sibani$^{3}$}
\affiliation{$^{1}$Department of Physics, Emory University, Atlanta, GA 30322,
USA~\\
$^{2}$Department of Earth Science and Engineering, Imperial College,
London SW7 2BP, United Kingdom~\\
$^{3}$Institut for Fysik Kemi og Farmaci, Syddansk Universitet, DK5230
Odense M, Denmark}
\begin{abstract}
Cooperative events requiring  anomalously
large fluctuations, are a defining characteristic for the onset
of glassy relaxation across many materials. 
The importance of such intermittent events has been noted
in systems as diverse as superconductors, metallic glasses, gels,
colloids, and granular piles. Here, we show that prohibiting the attainment
of new record-high energy fluctuations -- by explicitly imposing
a ``lid'' on the fluctuation spectrum -- impedes further relaxation
in the glassy phase. This lid allows us to directly measure the impact
of record events on the evolving system in extensive simulations of
aging in such vastly distinct glass formers as spin glasses and
 tapped granular piles.
 Interpreting our results in terms of a dynamics
of records succeeds in explaining the ubiquity of both, the logarithmic
decay of the energy and the memory effects encoded in the scaling
of two-time correlation functions of  aging systems. 
\end{abstract}
\maketitle

\section{Introduction\label{Intro}}
One of the simplest protocols to study the dynamics of non-equilibrium
systems consists of a hard quench applied 
when taking  them  instantly from very high to a low temperature. Of course,
in most ordinary materials (like a cup of coffee) heat is rapidly  expelled from the system      
so that the internal energy falls exponentially fast to the equilibrium
value that comports with its low temperature. However, when the quench
traverses a phase transition, the relaxation process becomes more
complicated. In particular, if a transition into a glassy phase is
passed~\cite{Struik78}, many out-of-equilibrium phenomena are observed,
often for a large range of time-scales~\cite{DH2011}. In fact, the
trajectory of the cooling system through configuration space for various
temperature protocols might serve as a ``microscope'' into the complexities
of the multimodal free-energy landscape of the glassy material~\cite{Bouchaud01}. 
The processes by which glassy systems scale barriers to evolve in their  landscape
has been, and still remains  a very active field of research~\cite{Amir2012,Lubchenko2017,Ros21}.

\begin{figure}
\hfill{}\includegraphics[bb=150bp 420bp 650bp 610bp,clip,width=1\columnwidth]{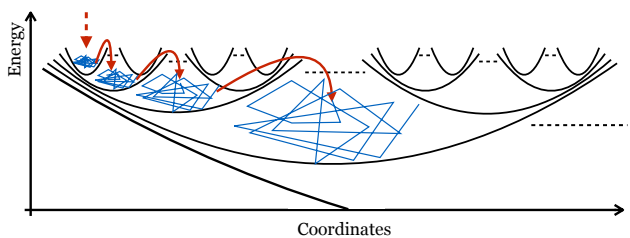}\hfill{}
\caption{\label{fig:landscape}Illustration of the high-dimensional energy
landscape traversed by a glass. On quenching (top arrow), it likely
is trapped in a metastable state (black basins) at higher energies,
vastly outnumbering those at lower energies. In the hierarchy~\cite{Sibani89,Charbonneau13},
higher-energy basins are shallower than lower ones. Thus, traversing
ever deeper basins (blue lines), ever larger energy barriers must
be surmounted (hooked arrows).}
\end{figure}

Remarkably, at least for the basic protocol of a hard quench, a large
class of glassy systems as diverse as spin glasses~\cite{Rodriguez03},
superconductors~\cite{Oliveira05}, gels~\cite{Parker03}, 
colloids~\cite{Yunker09,Kajiya13,Robe16}, and granular piles~\cite{GB20}
exhibit a ubiquitous behavior, irrespective of microscopic details.
Macroscopic observables evolve very slowly, essentially logarithmically,
and data for two-time correlation functions scales with the \emph{ratio}
of both times (``full aging''). Furthermore, it is found that the
aging dynamics is activated via intermittent, irreversible 
events~\cite{Bissig03,Buisson03,Sibani06a,Yunker09,Kajiya13}. These observations
can be unified via generic features, illustrated in Fig.~\ref{fig:landscape},
that are common to glassy energy landscapes~\cite{Heuer08,Robe16,Boettcher20c}
and has lead to models of relaxation on tree-structures~\cite{Sibani89,Fischer08}, which capture key features of real aging systems.
In spin glasses these features emerge because (1) a quench
leaves them  with many small domains at high energy and (2) enlarging
a domain requires overcoming barriers proportional to the size of
that domain~\cite{Fisher88,Shore92}.

In such a landscape, not only do its equilibrated domains grow exceedingly
slowly, in want of large fluctuations~\cite{Fisher88a}, but the glassy
system also \emph{must} achieve ever new records in its fluctuation
spectrum~\cite{Sibani93a} to maintain its relaxation. That records drive the 
dynamics (i.e., ``set the clock'') in a large class of complex systems, of which 
glassy materials are a subset, is referred to as Record Dynamics (RD)~\cite{Sibani21}. 
In the latter, quakes, the non-equilibrium events triggering the changes from a metastable 
state to the next are identified and shown to form a log-Poisson process~\cite{SJ13,Boettcher18b}. 
This means that the coarse-grained decelerating dynamics of the system at hand is log-time 
homogeneous and opens a way for a number of theoretical predictions. 
The rate of quakes and their  log-Poisson statistics have
been observed \emph{directly} in various glasses~\cite{Robe16,Robe18,Boettcher18b,Sibani18,GB20}. 
That records of the free energy, or some other fluctuating configuration space function,  actually trigger  
quakes is supported in RD  by analogies and by indirect evidence. The basic analogy is that records
in white noise, standing in for equilibrium fluctuations, are a log-Poisson process, independently of the
distribution from which the noise is drawn. The indirect evidence is that a relaxation on tree models of 
configuration space share key properties of aging systems~\cite{Sibani89,Fischer08}.

Here, we show how record 
events strongly influence the course of glassy   relaxation by 
 letting  a Maxwell demon intervene into the microscopic dynamics. 
The demon  explicitly imposes a \textquotedblleft lid\textquotedblright{}
on upward energy fluctuations. This eliminates the emergence of new high energy  records
and, at the same time, limits the system's ability to overcome energy barriers.
 Specifically,
we compare the behavior with and without a lid in extensive simulations
of both, an Ising spin glass in contact with a heat bath and a \emph{3d}
pile of grains, tapped upwards against gravity to activate
grains that dissipate through friction and inelastic collisions. 
In the spin glass  case, our  lid device  is shown to have very different 
consequence above and below the critical temperature. 
Energy relaxation toward lower  equilibrium values is  accelerated
above, but decelerated below $T_c$. In the granular pile,  the lid
nearly   stops the logarithmic decay of the 
potential energy..
\begin{figure}
\hfill{}\includegraphics[bb=0bp 40bp 785bp 618bp,clip,width=0.9\columnwidth]{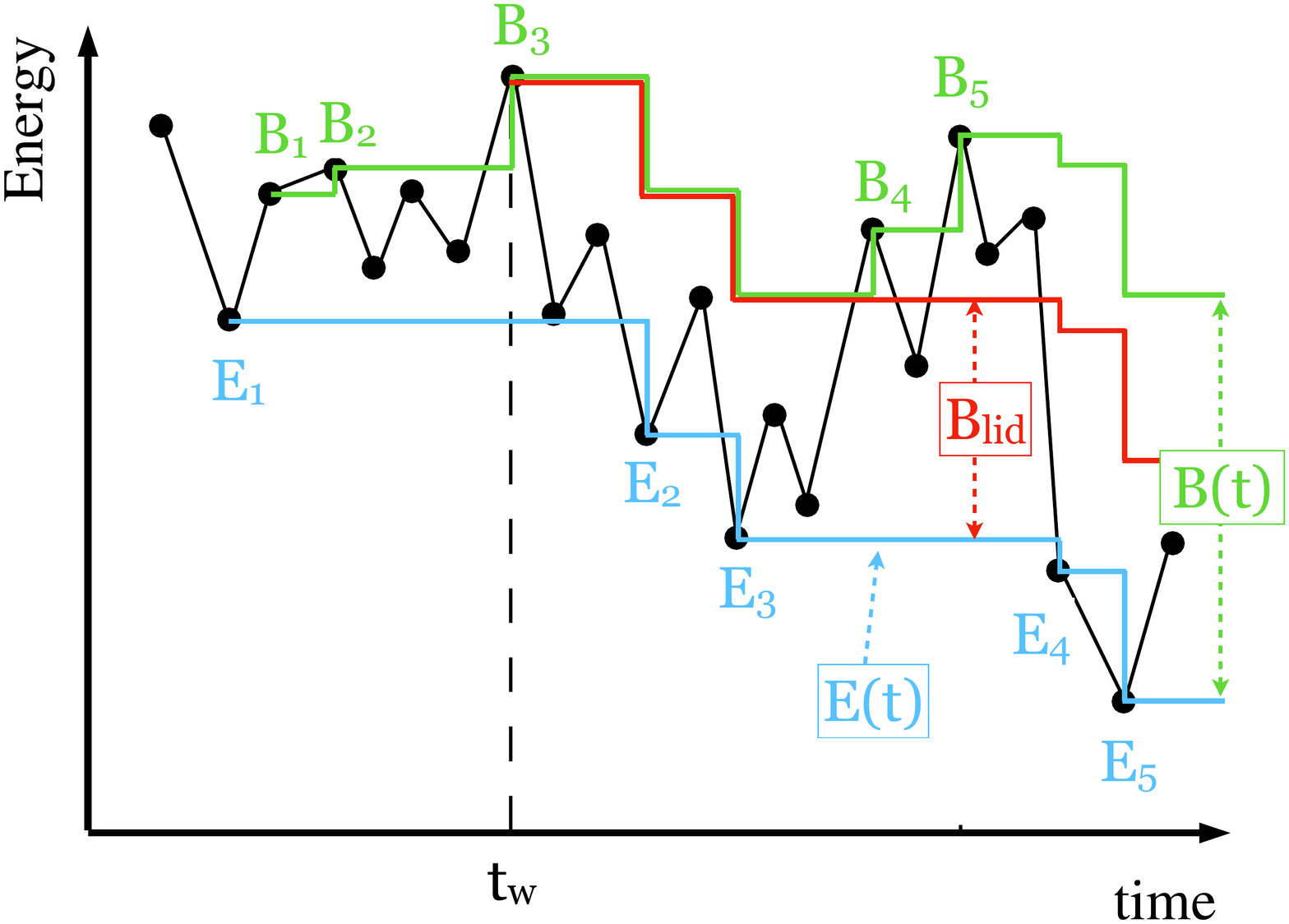}\hfill{}
\vspace{-0.25cm}
\caption{\label{fig:eTrace}Illustration for a trace of energy density $e(t)$
(black dots) during relaxation. We define the lowest energy encountered
up to time $t$, ${\rm E}(t)=\min_{0\protect\leq t^{\prime}\protect\leq t}\left\{ e\left(t^{\prime}\right)\right\} $
and the up-to-now highest ``barrier'' ${\rm B(t)}=\max_{0\protect\leq t^{\prime}\protect\leq t}\left\{ e\left(t^{\prime}\right)-{\rm E}\left(t^{\prime}\right)\right\} $
above ${\rm E}(t)$. These are staircase functions, with changes marked
by ${\rm E}_{i}$ and ${\rm B}_{j}$, that envelope the fluctuations
of $e(t)$ from below (blue) and above (green or red), respectively.
The red line marks a ``lid'' ${\rm B_{lid}}={\rm B}(t_{w})$ on
the barrier ``locked in'' at time $t_{w}$ (i.e., ${\rm B_{lid}}={\rm B}_{3}$
here) that curbs $e(t)$ for $t>t_{w}$. That is, updates like those
increasing $e(t)$ to ${\rm B}_{4}$ or ${\rm B}_{5}$ here, or to
any value above ${\rm E}(t)+{\rm B_{lid}}$, would be rejected with
such a lid.}
\end{figure}

\section{Quench Protocols, with or without a Lid\label{subsec:Quench-Protocols}}

In our aging simulations of the spin glass or the granular pile, we 
start in each case either with randomly assigned spins or with a 
randomly poured pile at $t_{0}=0$, respectively.
The ensuing relaxation process then corresponds to an aging protocol
for a hard (instantaneous) quench at $t=t_{0}$, in temperature from
$T=\infty$ to some finite $T$ for the Ising magnets, or in tapping
strength from infinitely hard to mild tapping. The ensuing energy trace $e(t)$
is illustrated in Fig.~\ref{fig:eTrace}, while examples of actual traces from our
experiments are shown below in Fig.~\ref{fig:EAtrace} for EA
or in Fig.~\ref{fig:taptrace} for the granular pile. 

As those traces of $e(t)$ show, the energy density decreases on average
while exhibiting a characteristic range of fluctuations. This (logarithmic) 
decline in average energy suggests that we need to assess the effect of fluctuations with
respect to an equally descending energy-scale. Note that this decline renders as tenuous the notion 
of aging as a return process (to some fixed, global energy level), as is inherent to  
the trap model description of aging~\cite{Bouchaud92,BaityJesi18}.
Instead, it is essential to assess records in these fluctuations~\cite{Dall03,Sibani18,Boettcher20c},
relative to the up-to-now lowest energy value encountered
at time $t$, ${\rm E}(t)=\min_{0\leq t^{\prime}\leq t}\left\{ e\left(t^{\prime}\right)\right\} $
and the ``barrier'' ${\rm B(t)}=\max_{0\leq t^{\prime}\leq t}\left\{ e\left(t^{\prime}\right)-{\rm E}\left(t^{\prime}\right)\right\} $,
i.e., the up-to-now highest energy attained \emph{relative} to the
most recent E. An energy trace $e(t)$ then maps into a random sequence
of symbols ${\rm E}_{i}$ and ${\rm B}_{j}$, as illustrated in Fig.~\ref{fig:eTrace}.
In the traces shown in Figs.~\ref{fig:EAtrace} or~\ref{fig:taptrace},
the values of ${\rm E}(t)$ and of ${\rm E}(t)+{\rm B}(t)$ are plotted
as envelopes below and above, respectively, to the fluctuating signal
$e(t)$. For the Ising magnets, the unobstructed fluctuations are
shown in green, while those shown in red have been curbed by a ``lid''. Although 
the fluctuations for the granular pile seem minute, they distribution is well-resolved~\cite{GB21}.

The manner in which the Maxwell demon imposes a lid on the 
fluctuations~\cite{Sibani99} is also illustrated 
in Fig.~\ref{fig:eTrace}: As ${\rm B}(t)$ represents
the span of regular (unhindered) fluctuations,  beginning
at some predetermined ``waiting time'' $t_{w}$ after the quench, the ``lid''
${\rm B_{lid}}={\rm B}(t_{w})$ rejects any update towards
exceeding ${\rm B_{lid}}$ for all $t>t_w$. For example, we chose $t_{w}=32$
for the Ising magnets and $t_{w}=100$ for the granular pile. Then, for all $t>t_{w}$
Monte Carlo updates of individual spins in the Ising magnets, or entire
taps for the granular pile, are rejected if they would increase at
any time to an energy of $e(t)\geq{\rm E}(t)+{\rm B_{lid}}$. Thus,
for such a lid no new barrier records ${\rm B}_{j}>{\rm B_{lid}}$
can ever be achieved! This is again illustrated in Fig.~\ref{fig:eTrace},
where the red envelope remains stuck at ${\rm B_{lid}}(={\rm B}_{3}$
in this illustration) while the green envelope -- without a lid --
achieves new barrier records ${\rm B}_{4},{\rm B}_{5},\ldots$. The
question addressed here is: Would the imposition of the
lid impede, or enhance, finding better minima, like ${\rm E}_{4}$
and ${\rm E}_{5}$ in this illustration?

\begin{figure}
\hfill{}\includegraphics[bb=0bp 20bp 730bp 600bp,clip,width=.97\columnwidth]{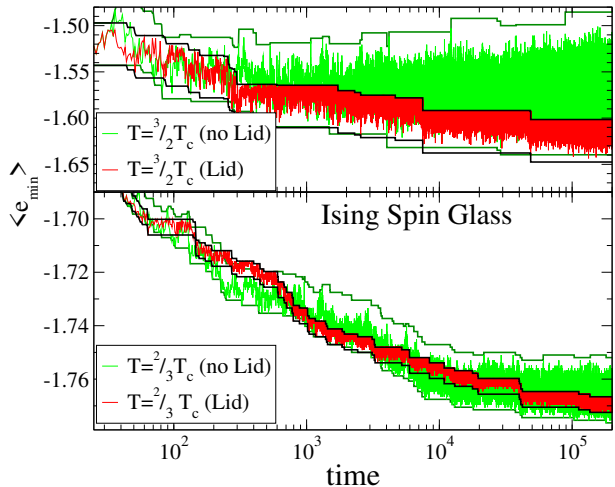}\hfill{}
\caption{\label{fig:EAtrace}Traces of the energy density $e(t)$ for the EA
spin glass after a quench at $t=0$ from $T=\infty$ to $T>T_{c}$
(top) or $T<T_{c}$ (bottom). Regular fluctuations with envelopes
are plotted in green, those using a lid are red with black envelopes.}
\end{figure}

\begin{figure}
\hfill{}\includegraphics[bb=155bp 80bp 610bp 610bp,clip,width=1\columnwidth]{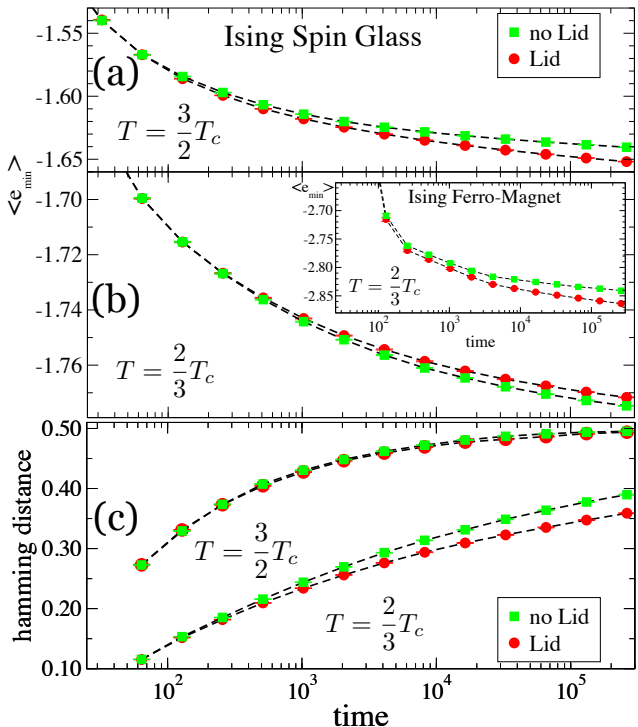}\hfill{}
\caption{\label{fig:lidplot}Plots for Monte Carlo simulations of the Ising
spin glass for $t>t_{w}=32$. Green squares refer to data obtained
with regular fluctuations, red circles to data with a lid ${\rm B_{lid}}$
imposed at time $t_{w}$, as defined in Fig.~\ref{fig:eTrace}.
Shown are the average minima $\left\langle {\rm E}(t)\right\rangle $
for energy fluctuations $e(t)$ (illustrated in Fig.~\ref{fig:EAtrace})
after a quench at $t=0$ from $T=\infty$ to $T>T_{c}$ (a) and $T<T_{c}$
(b). Inset in (b) shows the same at $T<T_{c}$ for the Ising ferromagnet.
In (c), the Hamming distances relative to the configuration found
at $t_{w}$ are shown for both protocols. All data was averaged over
3000 independent runs on lattices of size $N=16^{3}$, obtaining errors
smaller than symbol sizes.}
\end{figure}

\section{Spin Glass Fluctuations\label{SGFluc}}
The well-known Ising spin glass model on a \emph{3d} lattice due to
Edwards and Anderson (EA)~\cite{Edwards75} has become the very definition
of complexity and disorder, not only in physics but also in computer
science, engineering, and biology~\cite{MPV,Stein13}. We employ 
EA on hyper-cubic lattices of size $N=16^{3}$ with periodic boundary
conditions using the Hamiltonian 
\begin{equation}
H\left(\left\{ \sigma_{i}\right\} \right)=-\sum_{\left\langle i,j\right\rangle }J_{ij}\sigma_{i}\sigma_{j}.
\end{equation}
Bonds between neighboring spins $\left\langle i,j\right\rangle $
are drawn with equal probability to be $J_{ij}=\pm1$ and 1000 such
instances were generated randomly. As a reference, we have also studied 
purely ferromagnetic systems (FM) with only positive bonds, $J_{ij}\equiv J=1$, 
throughout. In each run, we have updated the system by the Metropolis method~\cite{Metropolis1953}
at a fixed temperature $T$, both for a higher ($T=\frac{3}{2}T_{c}$)
and for a lower temperature ($T=\frac{2}{3}T_{c}$), using $t_{{\rm max}}\approx3\,10^{5}$
system sweeps and tracked the instantaneous energy density of the
entire system, $e(t)$.  

We illustrate the typical behavior of $e(t)$
in Fig.~\ref{fig:EAtrace} with a single trace of the energy fluctuations
for times $t>t_{w}=32$ after the quench on a logarithmic time scale.
The wider fluctuations (green) represent the regular relaxation process,
while narrower fluctuations (red) result from the imposition of a
lid ${\rm B_{lid}}$ for $t>t_{w}$. As defined in Fig.~\ref{fig:eTrace},
the range of each signal is characterized by staircasing envelope-functions,
for ${\rm E}(t)$ below, and above by ${\rm E}(t)+{\rm B}(t)$ for
the regular fluctuations or ${\rm E}(t)+{\rm B_{lid}}$ for the fluctuations
curbed by a lid. For $T>T_{c}$, the regular signal soon relaxes to
some stationary value of $\left\langle e(t)\right\rangle $ around
which it fluctuates. In those cases, the evolution with the imposed lid
consistently skirts the lower edge of the regular fluctuations. In
turn, traces in the glassy regime ($T<T_{c}$ ) continue to decrease
on average while exhibiting a significantly narrower range of fluctuations.
Accordingly, the lid ${\rm B_{lid}}$ also attains a smaller value.
Here, the corresponding fluctuations struggle to achieve the same
set of minima, ${\rm E}(t)$, as the regular fluctuations.

The impression conveyed by those illustrative traces are borne out
by a systematic study, shown in Fig.~\ref{fig:lidplot}. Here, we
are not looking at the average of the fluctuations but measuring the energy
minima $\left\langle {\rm E}(t)\right\rangle $ that the system can
access with or without a lid on the range of energy barriers it is
allowed to cross. While the thermal average $\left\langle e(t)\right\rangle$ 
provides little information about the local structure of the energy landscape, 
it is these minima that signify reaching new basins of local stability. Definitely
for the high-temperature quench to $T>T_{c}$, but also for the quench
into the ferromagnetically order phase at $T<T_{c}$ (see inset),
the imposition of a lid actually results in achieving \emph{better}
energy minima than without. In contrast, for the glassy case in
Fig.~\ref{fig:lidplot}(b), the lid limits the quality of
energy minima that can be explored. These results are robust and only
vary gradually by varying $t_{w}$, or when using Gaussian distributed
bonds instead of $J_{ij}=\pm1$. The interpretation of these findings
is as follows: At high $T$, or even for ferromagnetic order, achieving
new records in the barriers that are overcome by fluctuations is \emph{immaterial}
for access to new energy minima; average-size fluctuations suffice
to traverse any part of the configuration space at the respective temperature.
In fact, the lid forces the Monte Carlo process to spent more time
at lower energy and, hence, \emph{increases} the probability of finding
new minima. In contrast, in the glassy case, prohibiting barrier crossings
via a lid curbs the exploration of new meta-stable basins for their
minima and more likely traps the system in the local basin. Hence,
traversing those barriers between basins via record fluctuations,
as stipulated for RD, proves essential for the glassy system to relax.

\begin{figure}
\hfill{}\includegraphics[bb=20bp 0bp 710bp 390bp,clip,width=1\columnwidth]{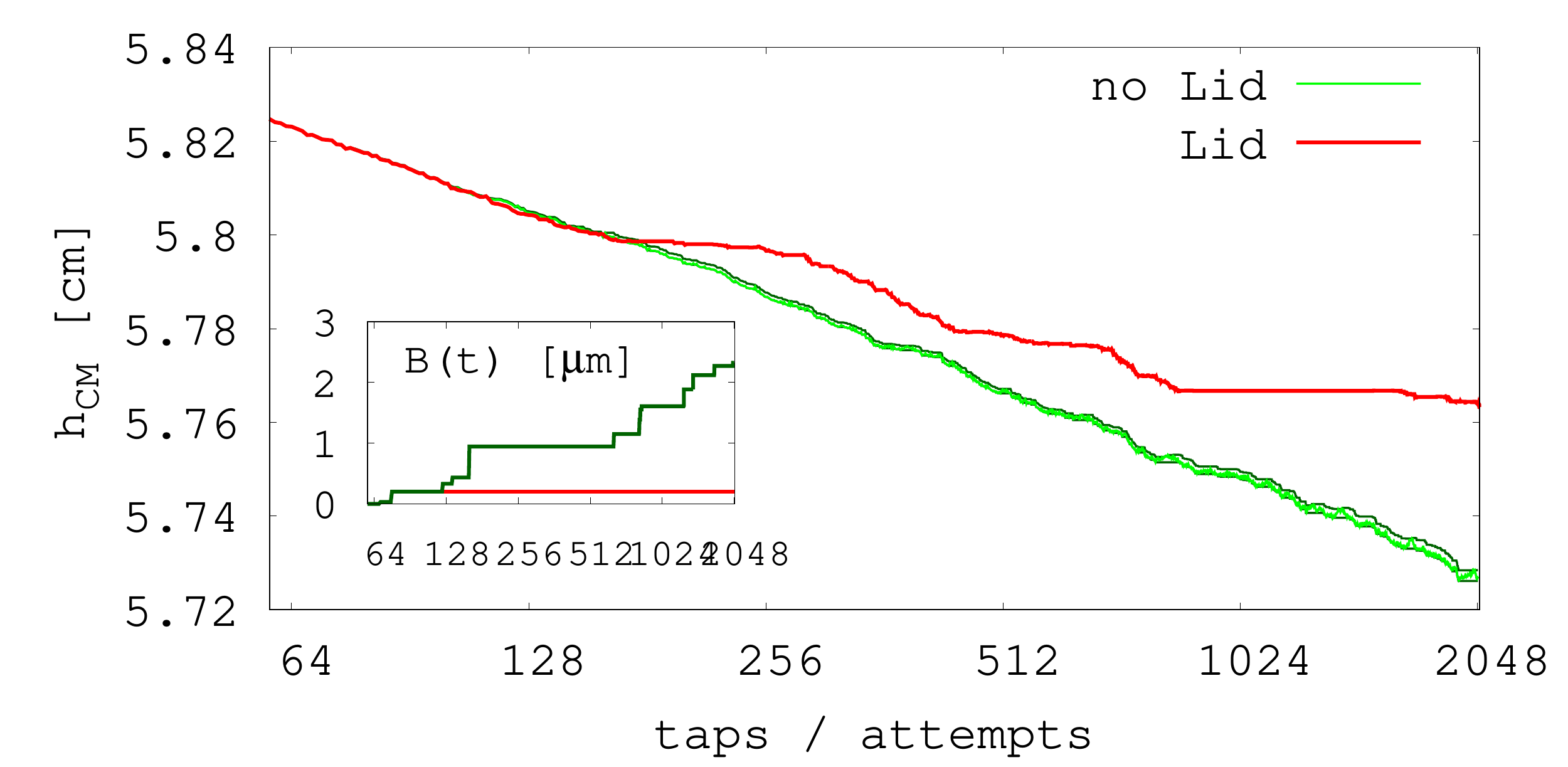}\hfill{}
\caption{\label{fig:taptrace}Trace for the potential energy (in units of center-of-mass
height $h_{{\rm CM}}$) with the number of taps in a granular pile
after a quench from strong to weak tapping, evolving regularly (green)
or with a lid ${\rm B_{lid}}$ imposed after $t_{w}=100$ taps (red).
With that lid, the evolution of the system stalls for increasingly
extended periods. Following the definition of ${\rm E}(t)$ and ${\rm B}(t)$
in Fig.~\ref{fig:eTrace}, lower and upper envelopes are shown for
the regular fluctuations, as in Fig.~\ref{fig:EAtrace}. \emph{Inset}:
Logarithmic progression in record barrier size ${\rm B}(t)$ (green)
and height of the lid ${\rm B_{lid}}$ (red). }
\end{figure}

\begin{figure}
\hfill{}\includegraphics[bb=20bp 0bp 710bp 554bp,clip,width=1\columnwidth]{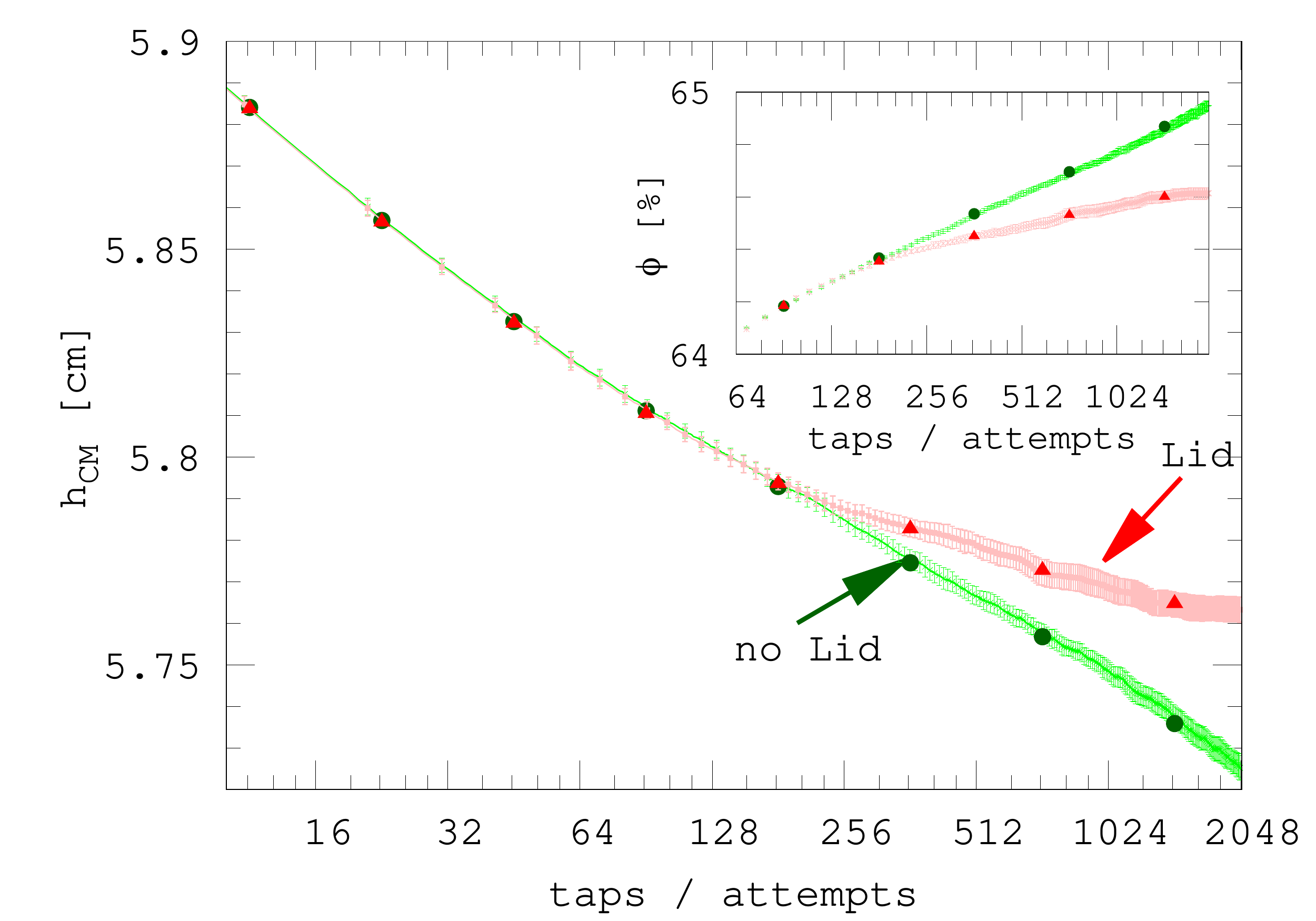}\hfill{}
\vspace{-0.2cm}
\caption{\label{fig:TapLid}Plot of the potential energy (in units of the center-of-mass
height of the pile, $h_{{\rm CM}}$) after a quench at $t_{0}=0$
for $t_{{\rm max}}=2048$ attempted taps, without (green data) and
with a lid ${\rm B_{lid}}$ imposed at $t_{w}=100$ (red data). Symbols
represent a log-binned average of the respective data that is itself
averaged after each tap over ten independent runs, like that shown
in Fig.~\ref{fig:taptrace}. The inset demonstrates the logarithmic
increase in the density of the pile during the same runs (without
lid in green), and the density stalling with a lid (red).}
\end{figure}

\section{Fluctuations in Granular Piles\label{subsec:Simulations-of-Granular}}

The wide scope of what we found for spin glasses is demonstrated in 
Figs.~\ref{fig:taptrace}-\ref{fig:TapLid}, where we study aging in a tapped granular 
pile using molecular dynamics simulations (MD).  To that end, we employ
the implementation recently presented in Refs.~\citenum{GB20,GB21}.
It is inspired by the experimental setup used for the Chicago experiment
\cite{Nowak98}. Our three-dimensional pile consists of $N=60,000$
bidisperse spheres, in equal proportion of 1 or 1.02mm in diameter,
contained in a vertical cylinder of 2.4cm diameter. The initial column
height, obtained through simple pouring, is about 11.5cm. The (minute)
bidispersity was introduced in order to prevent crystallization and
the particle diameter ratio was chosen to avoid segregation. We use
MD as implemented in the soft spheres model provided by the 
\emph{LIGGGHTS} open source software~\cite{goniva2012influence}, 
i.e.,  \emph{LAMMPS} improved for general granular and granular 
heat transfer simulations. Within \emph{LIGGGHTS},
we apply a Hertz model~\cite{brilliantov1996model} for the particle--particle
and particle--wall contact forces (with a Young\textquoteright s
modulus of $Y=6\times10^{6}{\rm Nm^{-2}}$, a Poisson ration of 0.3,
a restitution coefficient of 0.5, and a friction coefficient of $\mu=0.5$).
Particle densities equal $\rho=2,500{\rm kg/m^{3}}$.

Tapping simulations conducted entirely with MD evolve deterministically.
For the Maxwell demon to be able to reject a tap, we need each tap to contain a systemic
random process that would lead to alternative (randomized) outcomes
for a renewed tap of any given static configuration. In order to introduce
stochasticity here, instead of deterministically pushing the container 
floor and walls upward, we initiate each tap via a perturbation that
consists of a homogeneous vertical dilatation. It is obtained by multiplying
the $z$-coordinate of each particle in the system by a random amount
$\delta$ that is uniformly distributed in the range $1.003461539<\delta<1.004230769$.
This perturbation is followed by a MD deposition under the action
of gravity, until all kinetic energy is dissipated. Only the mechanically
static configurations at the end of each tap process are considered
in our measurements of the total gravitational potential energy of
the system. The system will be considered mechanically static when
the kinetic energy of the pile falls below $1.6\times10^{-8}{\rm J}$.
This cutoff value was deemed sufficient after examination of the system
dynamics and its kinetic energy evolution. The use of this homogeneous
vertical expansion to mimic the effect of a tap has been proposed
before for a simpler granular model with ballistic deposition~\cite{Philippe01},
which has proven to show good agreement in reproducing the dynamics
of the compaction as a function of the perturbation intensity~\cite{Philippe01,pugnaloni2008nonmonotonic}
and the signature effects of aging behavior.

We evolve ten instances of the pile after a quench at $t_{0}=0$ for
$t_{{\rm max}}=2048$ attempted taps, which is our unit of time in
this process. We measure the density of the pile in terms of the average
local volume fraction $\phi$ occupied by the particles, which is
obtaining from their Voronoi volume (using Voro++~\cite{Rycroft2009})
for all particles that are two particle diameters off the container
walls and its surface. The internal energy $e(t)$ here is presented
in terms of the center-of-mass height of the pile, $h_{{\rm CM}}=\left\langle h\right\rangle $,
where the total potential energy is $Nmgh_{{\rm CM}}$, with constants
for gravity $g$ and the mass of the pile, $Nm$. 

In Fig.~\ref{fig:taptrace}, we show a trace of the internal (gravitational)
energy for a single instance, without and with a lid ${\rm B_{lid}}$ imposed
after $t_{w}=100$ taps, similar to Fig.~\ref{fig:EAtrace}. Fig.~\ref{fig:TapLid}
shows the average of that energy over all ten instances, decreasing logarithmically
without a lid but stalling when the lid is imposed, similar to Fig.~\ref{fig:lidplot}.
In the inset of Fig.~\ref{fig:TapLid}, we demonstrate that a sequence
of those taps following a quench reproduces the logarithmic density
increase (shown as volume fraction $\phi$, in green) that is expected
from experiments~\cite{Nowak98} and that is consistent with a 
glass~\cite{Richard05,GB20,GB21}. Although there was no lid on the density fluctuation
itself, the lid on the energy fluctuations nonetheless stalls the
relaxation in density, also shown in that inset (red).

\section{Discussion\label{Discussion}}
We note that our  global
energy record  are  merely a stand-in for  irreversible events that
occur locally. In many contexts these are harder to identify and control
microscopically but certainly involve an additional entropic element
\footnote{These events are analyzed in Ref.~\citenum{Sibani18}
for spin glasses, and in Ref.~\citenum{GB20} for the granular pile.}.
For this reason,, the lid we define likely does not completely trap the
glassy system within a basin of attraction. The record barrier that the system overcomes
in any particular relaxation process is not necessarily in the absolute
lowest-energy path available to leave the basin. The latter is just a more probable
one to follow. In other words,  a purely energetic lid entirely ignores the entropic component 
of the free-energy barrier that  the system eventually finds after the
lid confines it long enough within a given basin. Thus, the system
ultimately escapes anyway and finds better energy minima, albeit much later.

We can directly assess the entropic aspects of the different relaxation
processes in the Ising magnets by also considering the Hamming distances
between the spin configurations at any later minima ${\rm E}(t)$
relative to the latest minimum encountered just before $t_{w}$, ${\rm E}(t_{w})$.
The Hamming distance here counts the number of Ising spins between
these configurations that are no longer in the same orientation, divided
by the total number of spins, $N$. Thus, identical configurations
have zero distance while each differing spin adds $1/N$. A distance
of $\frac{1}{2}$ for two configurations makes them maximally uncorrelated,
as those would share half of their spins purely at random. As the
bottom panel of Fig.~\ref{fig:lidplot} shows, at high temperature
$T>T_{c}$, the relaxation process completely decorrelates, independent
of the lid. Even at low temperature $T<T_{c}$ for a ferromagnet,
the Hamming distance is indistinguishable between having a lid or
not (not shown). However, we observe quite a different behavior for
the glassy case: Although the difference in the value of the minima
explored with or without lid might seem small, it certainly signals
a profound constraint on the part of configuration space that a glassy
system is allowed to explore when a lid is imposed! At the latest
time simulated here, the Hamming distance for new minima explored
by the process with a lid lags behind that of the regular exploration
by about $0.05$. That is, the range of exploration within the configuration
space is smaller by $\sim0.05N\approx200$ spins when record barrier
crossings are curbed with the lid, corresponding to a reduced volume
of some $2^{200}\approx10^{60}$ configurations that remain out of
range. Of course, only few of those get visited even without the lid,
and even fewer are novel minima, yet, those numbers indicate the enormous
consequences extremal fluctuations, or the lack thereof, entail in glassy
relaxation.

\section{Conclusion\label{Conclusion}}
We have demonstrated that, in glassy dynamics,
extremal upward energy fluctuations
are essential for  how extremal low energy states are reached
 while  the system descends through its
complex, multimodal energy landscape toward equilibrium.
 While
fluctuations in non-glassy systems also exhibit large excursions,
these remain inconsequential, since typical fluctuations are sufficient
for relaxation. Thus, any physical description of glassy  relaxation
 will have to account for those rare, irreversible
and record-sized events that drive the observed intermittent 
behavior~\cite{Bissig03,Buisson03,Sibani06a,Parker03,Yunker09,Kajiya13} and
manifest themselves across many 
materials~\cite{Rodriguez03,Oliveira05,Parker03,Yunker09,Kajiya13,Robe16,GB20},
irrespective of any microscopic details, as especially the example
of the (a-thermal) granular pile clearly demonstrates. These insights not
only pertain to aging behavior in physical systems~\cite{roth2016polymer},
but likely also affect the behavior of local search heuristics in
the optimization of hard combinatorial problems~\cite{Boettcher01a,BoSi}
and many other applications of glassy dynamics~\cite{Stein13}. 

\acknowledgements
Simulations of the granular pile were performed at the Imperial College
Research Computing Service (see DOI: 10.14469/hpc/2232). 

\bibliographystyle{apsrev4-2}
\bibliography{/Users/sboettc/Boettcher}

\end{document}